\documentclass[12pt]{article}
\usepackage{amssymb}
\def\inh{\vskip 0.075truein \noindent\hangindent=12 pt \hangafter=1}

\usepackage{amsthm}\theoremstyle{remark}

\newcommand{\bte}{\begin{quote}\begin{theorem}}
\newcommand{\ete}[1]{\label{#1}\end{theorem}\end{quote}}
\newcommand{\bcom}{\begin{quote}\end{quote}}
\newcommand{\bex}{\begin{quote}\begin{example}}
\newcommand{\eex}[1]{\label{#1}\end{example}\end{quote}}
\newcommand{\bcon}{\begin{quote}\begin{conclusion}}
\newcommand{\econ}[1]{\label{#1}\end{conclusion}\end{quote}}
\newcommand{\bdefi}{\begin{quote}\begin{definition}}
\newcommand{\edefi}[1]{\label{#1}\end{definition}\end{quote}}

\newcommand{\blem}{\begin{quote}\begin{lemma}}
\newcommand{\elem}[1]{\label{#1}\end{lemma}\end{quote}}

\newcommand{\bpr}{\begin{quote}\begin{problem}}
\newcommand{\epr}[1]{\label{#1}\end{problem}\end{quote}}

\newcommand{\f}{\frac}
\newcommand{\p}{\partial}
\newcommand{\n}{\nonumber \\}

\newcommand{\beq}{\begin{eqnarray}}
\newcommand{\eeq}[1]{\label{#1}\end{eqnarray}}
\newcommand\eq[1]{(\ref{#1})}
\newcommand{\bfi}{\begin{figure}[24]}
\newcommand{\efi}[1]{\caption{\label{#1}}\end{figure}}
\newcommand\fig[1]{Fig.~\ref{#1}}
\newcommand{\res}{respectively}
\newcommand\gl{\left}
\newcommand\gr{\right}

\newcommand{\CA}{{\cal A}}

\newcommand{\CE}{{\cal E}}

\newcommand{\Ga}{\alpha}
\newcommand{\Gb}{\beta}

\newcommand{\Gf}{\phi}

\newcommand{\Gg}{\gamma}

\newcommand{\Gk}{\varkappa}
\newcommand{\Gl}{\lambda}

\newcommand{\Gm}{\mu}

\newcommand{\Gr}{\varrho}

\newcommand{\Go}{\omega}

\newcommand{\GO}{\Omega}

\newcommand\D{\,\mathrm{d}}

\newcommand{\bexe}{\begin{quote}\begin{exercise}\inh}
\newcommand{\eexe}[1]{\label{#1}\end{exercise}\end{quote}}

\usepackage{graphics,graphicx}
\usepackage{color}
\begin{document}
{
\title{
Coupled mode parametric resonance in a vibrating screen model
}
}

\author{Leonid I. Slepyan$^{a*}$ and Victor I. Slepyan$^b$}
\date{$^a${\em School of Mechanical Engineering, Tel Aviv University\\
Ramat Aviv 69978 Israel} \\
$^b${\em Loginov $\&$ Partners Mining Company\\ 2v Korolyova Avenue 03134 Kiev, Ukraine}}

\maketitle

\vspace{5mm}\noindent
{\bf Abstract}

\vspace{3mm}
\noindent
We consider a simple dynamic model of the vibrating screen operating in the parametric resonance (PR) mode. This model was used in the course of  designing and setting of such a screen in LPMC. The PR-based screen compares favorably with conventional types of such machines, where the transverse oscillations are excited directly. It is characterized by larger values of the amplitude and by insensitivity to damping in a rather wide range. The model represents an initially strained system of two equal masses connected by a linearly elastic string. Self-equilibrated, longitudinal, harmonic forces act on the masses. Under certain conditions this results in transverse, finite-amplitude oscillations of the string. The problem is reduced to a system of two ordinary differential equations coupled by the geometric nonlinearity. Damping in both the transverse and longitudinal oscillations is taken into account. Free and forced oscillations of this mass-string system are examined analytically and numerically. The energy exchange between the longitudinal and transverse modes of free oscillations is demonstrated. An exact analytical solution is found for the forced oscillations, where the coupling plays the role of a stabilizer. In a more general case, the harmonic analysis is used with neglect of the higher harmonics. Explicit expressions for all parameters of the steady nonlinear oscillations are determined. The domains are found where the analytically obtained steady oscillation regimes are stable. Over the frequency ranges, where the steady oscillations exist, a perfect correspondence is found between the amplitudes obtained analytically and numerically. Illustrations based on the analytical and numerical simulations are presented.

\vspace{3mm}

\vspace{3mm}
$^*${\em Corresponding author: Leonid I. Slepyan,  leonid@eng.tau.ac.il, tel. +972 544 897 220.}

\section{Introduction}
\begin{figure}[!ht]

\vspace{-85mm}

\centering
\vspace*{10mm} \hspace{-15mm}
 \rotatebox{0}{\resizebox{!}{20cm}{%
\includegraphics [scale=0.25]{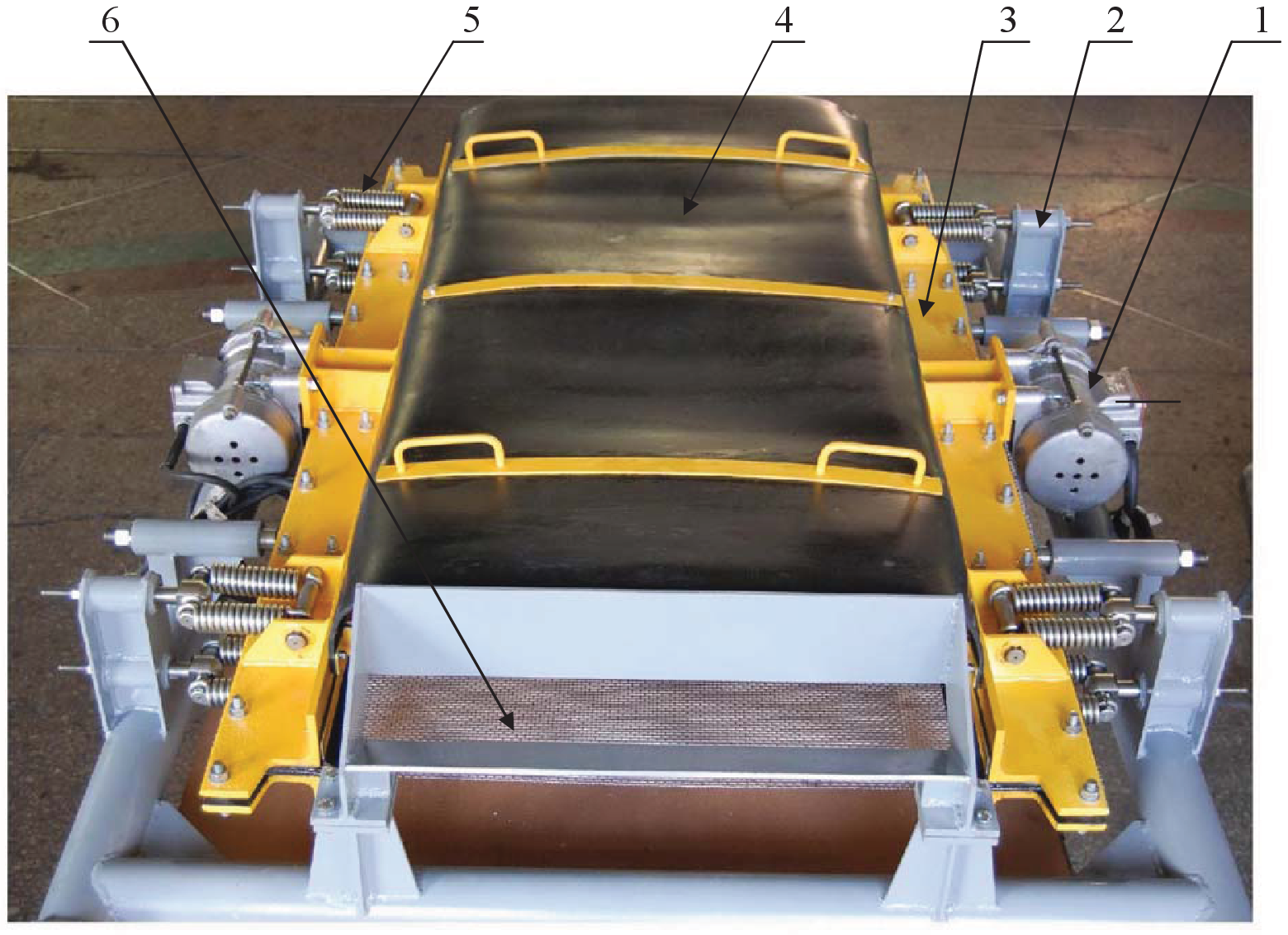}}}

\vspace{-15mm}
 \label{v-s}
\end{figure}

In this paper, parametric resonance (PR) is considered in a system related to the PR-based vibrating screen, \fig{v-s}.

\begin{figure}[!ht]

\vspace{-50mm}
\centering
\vspace*{-50mm} \rotatebox{0}{\resizebox{!}{9.5cm}{%
\includegraphics [scale=0.25]{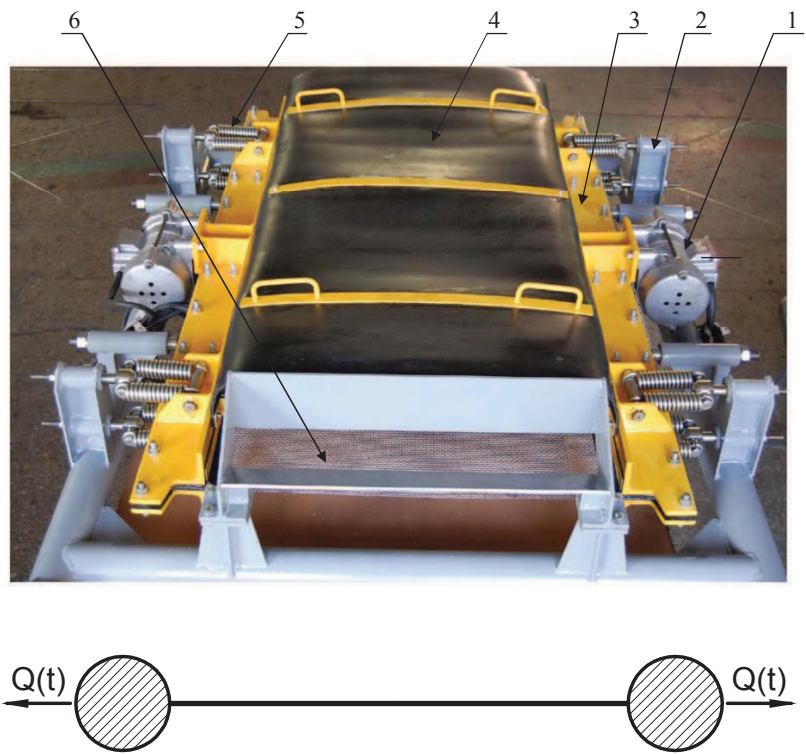}}}

\vspace{-18mm}
\caption{The vibrating screen and its simplest model. In the screen photo:  the vibrator (1), the base (2), the beams fastening the sieve (3), the lid of the vibrating screen (4), the side springs (5) and the sieve (6) (it is mainly under the lid). In the model: the end masses can move synchronously in opposite horizontal directions, while the string can oscillate laterally. These two modes of oscillations are coupled because the tensile force depends on both the longitudinal displacements of the masses and the transverse displacements of the string (the latter dependence is nonlinear).}
\label{model}
\end{figure}

\vspace{0mm}
The idea to create such a machine came to us in 2007 while discussing drawbacks of the existing types of the screens. In 2009, the patent was issued on the excitation method of the screen and the corresponding structure of the latter [1]. At that time, the nonlinear dynamics of such a machine had been numerically simulated and the first PR-based screen was built in Loginov and Partner Mining Company (Kiev, Ukraine). The PR-based screen compares favorably with conventional types of such machines, where the transverse oscillations are excited directly. It is characterized by larger values of the amplitudes and by insensitivity to the dissipation level in a rather wide range of the viscosity.

The model considered here or a similar one can also be found useful in some other PR applications. At the same time, the stable operation of a PR-based machine assumes the proper design and setup, which can be achieved on the basis of the mathematical analysis of its dynamics. We now consider the problem analytically and numerically. The numerical simulations serve for the refinement of the domains, where the parametric oscillations are excited and where the analytically obtained steady oscillation regimes are stable. Both the analytical and numerical simulations are used for the illustration and verification of the results.

The problem is described by a system of two coupled nonlinear equations. We find an exact solution of these equations, which exists in the case of no damping associated with the transverse oscillations. This solution corresponds to an invariable tensile force. The equations appear uncoupled and linear; however, the solution is bounded and uniquely defined by the fact that the nonlinearity in this established regime is zero. In a domain of the problem parameters, it is stable due to the presence of the nonlinearity `in the background'. In this case, the coupling plays the role of a stabiliser. It is demonstrated numerically how the transient regime approaches in time this established one defined analytically.

Next, we consider a more general, truly nonlinear regime. We use the harmonic analysis with the higher harmonics neglected. In the considered case, the latter simplification has virtually no effect on the results obtained. Explicit expressions are found for the amplitudes of longitudinal and transverse oscillations as functions of the external force amplitude and frequency. It is remarkable that in the case of the resonant excitation, where the external force frequency coincide with the frequency of the free longitudinal oscillations, the amplitudes are independent of the viscosity. In this regime, the nonlinearity bounds the amplitudes and the damping provides the stability.

Along with the nonlinear problem, the boundaries of the PR domain in the frequency-amplitude plane are determined based on the linear formulation. The PR arises in the nonlinear problem practically in the frequency region predicted by the linear analysis, slightly shifted towards the higher frequencies. It is shown that there is a sub region in this PR region, where the analytically obtained steady oscillation regime is stable being reproduced in numerical simulations with a high accuracy. The steady PR regime can exist in a structure dependent range of the frequency with a moderate nonzero damping level. The transverse oscillations, regular or irregular, abruptly decaying on the boundaries do not exist outside of the PR region. Amplitude-frequency characteristics and some PR regimes realizations are presented.

Before the forced PR regime we consider free oscillations of the perfect structure. The periodic energy exchange between the longitudinal and transverse modes of oscillations is demonstrated. It is shown, in particular, that the period increases as the energy decreases. Note that this regime is, in a sense, similar to that for the spring pendulum system (Vitt and Gorelik [2], Lai [3], Gaponov-Grekhov and Rabinovich [4]).

While in the past the parametric resonance has been mainly considered as an undesirable phenomenon, some attempts were made to employ it to obtain a greater response to excitation.  Related problems were studied mainly (but not only) in the application to micro devices (see, e.g., Baskaran and Turner [5], Roads et al [6], Krylov [7], Krylov et al [8-10], Fey at al [11], Fossen and Nijmeijer [12], Plat and Busher [13] and the references therein).

\section{The model}
Recall that the simplest model of the PR-based vibrating screen represents an initially stretched system of two equal masses connected by an elastic string, \fig{v-s}. The end masses, which can move horizontally, are also connected with a rigid frame by the side springs (as is discussed below their action can be considered as an invariable force). The longitudinal oscillations are excited by harmonic external forces acting synchronically on the left and the right masses in opposite horizontal directions. Under certain conditions this results in transverse oscillations of the string. In this study, the treated granular material action is reflected by linear viscosity.

We use the following notations: $2l, \Gr$ and $k/2$ are the string length, mass per unit length and stiffness, \res, $M$ is the end mass value, $\Gk$ is the side spring stiffness, $T_0$ is the initial tensile force in the string, $u(t)$ is the displacement of the right end mass ($-u(t)$ corresponds to the left one), $v(x,t)$ is the transverse displacement of the string ($-l<x<l$), $w(t)=v(0,t)$, $\Gb$ and $\Ga$ are the viscosity numbers associated with the longitudinal and transverse oscillations, \res.  The total tensile and external forces are denoted as
 \beq T(t)=T_0+ T_1(t)\,,~~~ Q(t)=T_0 +Q\cos \Go t\,. \eeq{0}

It is assumed that
 \beq \Gr l \ll M\,,~~~T_0\ll kl\,,~~~\Gk\ll k\,,~~~w\ll 2l\,,~~~\Go l \ll c_L=\sqrt{\f{kl}{\Gr}} \,, \eeq{001}
where $c_L$ is the longitudinal wave speed in the string. No limitation is imposed on the external harmonic forces, whereas the string is assumed to resist only to the positive tension, $T(t)\ge 0$ (we do not impose this condition in the formulation but check whether it is satisfied). Only the main mode of oscillation is considered. The assumptions \eq{001} allow us to neglect the string density in the equation of the longitudinal oscillations, to neglect the variation of the side spring action, to consider the tensile force to be independent of the coordinate, to consider the action of the string on the end mass as directed horizontally and to simplify the expression of nonlinearity caused by the string bending.

So, if there is no dissipation the frequencies of the longitudinal and transverse small-amplitude free oscillations are
 \beq \GO_L=  \sqrt{\f{k}{M}}\,,~~~\GO_T= \f{\pi}{2l}c_T\,,~~~c_T=\sqrt{\f{T_0}{\Gr}}\,,\eeq{sm2}
where $c_T$ is the transverse wave speed in the string.

If the longitudinal excitation frequency, $\Go$, is close enough to $2\GO_T$ and its amplitude, $Q$, is large enough, the parametric resonance arises under which the amplitude of the transverse oscillations is bounded by the geometric nonlinearity. We have the following nonlinear dynamic equations with respect to the longitudinal and transverse oscillations
 \beq M\f{\D^2 u(t)}{\D t^2} +\Gb\f{\D u(t)}{\D t}+ T_1(t)=Q\cos\Go t\,,\n
 \Gr\f{\p^2 v(x,t)}{\p t^2}+\Ga\f{\p v(x,t)}{\p t}  - T(t)\f{\p^2 v(x,t)}{\p x^2}= 0\,.\eeq{sm3}
These equations are coupled by the tensile force $T_1(t)$ which depends on both the longitudinal displacements of the masses, $ u(t)=u(t,l)=-u(t,-l)$, and the transverse displacement of the string, $v(x,t)$
 \beq T_1(t)=k\gl[u(t) +\int_0^l \sqrt{1+(v'(x,t))^2} \D x -l\gr]\,,\eeq{sm4}
where $\Ga$ and $\Gb$ are the viscosity numbers, $v'(x,t)=\p v(x,t)/\p x$ and expression \eq{sm4} is valid if it defines a nonnegative tensile force; otherwise, $T(t)=0$.
In this study, we consider the problem assuming that this condition is satisfied; then we determine domains where this is so.

Clearly, the left side of the second equation in \eq{sm3} with the boundary conditions, $v(\pm l,t)=0$, admits the variables separation as
 \beq v(x,t) = w(t)\cos \f{\pi x}{2l}\,.\eeq{sn5}
Using this expression and the condition concerning the amplitude of the transverse oscillations in \eq{001}, the expression for the tensile force \eq{sm4} can be reduces to
 \beq T_1(t)= k\gl(u(t) + \f{\pi^2}{16l} w^2(t)\gr)\,.\eeq{sm6}
The system of the dynamic equations becomes
 \beq M\f{\D^2 u(t)}{\D t^2} +\Gb\f{\D u(t)}{\D t}+ T_1(t)=Q\cos\Go t\,,\n
 \Gr \f{\D^2 w(t)}{\D t^2}+\Ga\f{\D w(t)}{\D t}  +\f{\pi^2}{4l^2}T(t)w(t)=0\,.\eeq{sm7}
This system is the base of the below considerations. Note that this model and some its generalizations were used in the numerical simulations while designing and setting of the PR vibrating screen. Being useful for the machine structure and setting pre-select it is also most suitable for the analytical study.

\subsection{Energy relations}
The sum of kinetic and potential energies of the system is
 \beq  \CE = M[\dot{u}(t)]^2 + \f{1}{2}\Gr l[\dot{w}(t)]^2 +\f{1}{k}T(t)^2 - 2T_0u(t)\,.\eeq{ex01}
With accuracy of a constant it is reduced to
 \beq  \CE = M[\dot{u}(t)]^2 + \f{1}{2}\Gr l[\dot{w}(t)]^2 +\f{1}{k}T_1(t)^2 +T_0\f{\pi^2}{8l}w^2(t)\,.\eeq{ex1}
We consider the energy consisting of longitudinal and transverse parts as follows
 \beq \CE=\CE_u+\CE_w\,,\n
 \CE_u = M\dot{u}^2(t)+\f{1}{k}T_1^2(t)~~~(\mbox{longitudinal part})\,,\n
 \CE_w=\f{1}{2}\Gr l\dot{w}^2(t) +\f{\pi^2}{8l}T_0 w^2(t)~~~(\mbox{ transverse part})\,.\eeq{ex2}
From this and the dynamic equations \eq{sm7} we find the energy rates
 \beq \dot{\CE}_u= 2(M\ddot{u}(t) +T_1)\dot{u}(t) +\f{\pi^2}{8l}T_1\f{\D w^2}{\D t}=2Q\dot{u}(t)\cos\Go t - 2\Gb[\dot{u}(t)]^2 +\f{\pi^2}{8l}T_1\f{\D w^2(t)}{\D t}\,,\n
 \dot{\CE}_w = \Gr l\ddot{w}(t)\dot{w}(t) +\f{\pi^2}{8l}T_0\f{\D w^2}{\D t} = - \Ga l[\dot{w}(t)]^2 -\f{\pi^2}{8l}T_1\f{\D w^2(t)}{\D t}\,.\eeq{ex3a}
Recall that the nonlinear term is responsible for the energy exchange between the oscillation modes.

\subsection{Free oscillations}
\begin{figure}[!ht]

\centering
\vspace*{10mm} \rotatebox{0}{\resizebox{!}{5
cm}{%
\includegraphics [scale=0.25]{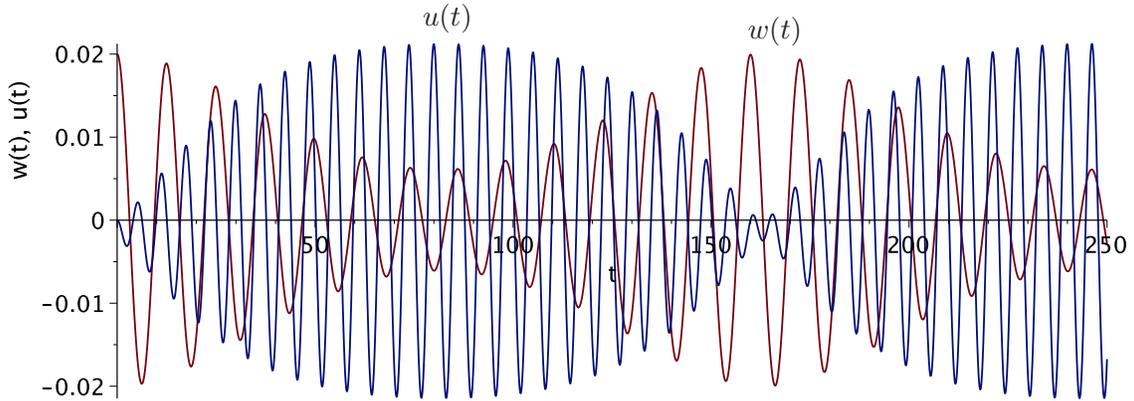}}}
\small \put(-147,140) {$w(t)$}
\small \put(-270,145) {$u(t)$}
 \caption{ The transverse and longitudinal free oscillations,  $w(t)$[m] and $u(t) [10^{-1}$m].}
    \label{f2}
\end{figure}

\begin{figure}[!ht]

\centering
\vspace*{-5mm} \rotatebox{0}{\resizebox{!}{4.5cm}{%
\includegraphics [scale=0.25]{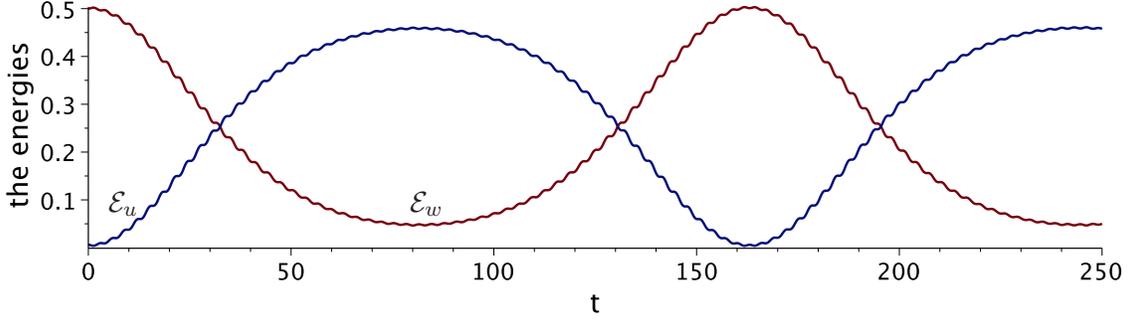}}}
\small \put(-273,43) {$\CE_w$}
\small \put(-387,43) {$\CE_u$}
 \caption{. The energy exchange between the longitudinal and transverse free oscillations, $\CE_{u,w}$ [$10^{-3}$ Nm]. The `roughness' of the curves is a consequence of numerous zeros of $\dot{\CE}_{u,w}$.}
    \label{f3}
\end{figure}

For free oscillations, $Q=\Ga=\Gb=0$, we have
 \beq \dot{\CE}_u=-\dot{\CE}_w  = \f{\pi^2}{8l}T_1\f{\D w^2(t)}{\D t}\,.\eeq{ex3}
An example of the free oscillations and the energy exchange is shown in \fig{f2} and \fig{f3}. The displacements, $w(t)$[m], $u(t)[10^{-1}$m] and the energies, $\CE_{u,w} [10^{-3}$Nm] are calculated for the system with $M$=100 kg, $\Gr$=10 kg/m, $T_0$=1 N, $k$=100 N/m, $l$=1 m (these units and a second as the time-unit are used here and below), under the initial conditions $w(0)$=0.02 m, $ u(0)=\dot{w}(0)=\dot{u}(0)=0$.

Note that, in accordance with \eq{ex3}, the energy exchange period increases to infinity as the energy decreases to zero. Indeed, for given dynamic parameters of the system the frequency weakly depends on the energy, that is, on the oscillation amplitude; hence the derivative, $\D w^2(t)/\D t$ is of the same order as $w^2$ and the energy $\CE_w$, \eq{ex2}. The variable part of the tensile force, $T_1$, also decreases to zero as the energy. Thus, the derivative, $\dot{\CE}_w$, decreases faster than the energy itself, that results in the increase of the energy exchange period. Numerical simulations support the guess that the period is asymptotically inversely proportional to the energy.

\section{An exact solution}
It is interesting that there is an elementary exact solution of the system \eq{sm7}. It exists in the case of the resonance excitation, $\Go=2\GO_T$, of the model with no damping associated with the lateral oscillations, $\Ga=0$. In this solution, the tensile force is invariable:
 \beq T(t) \equiv T_0\,,~~\mbox{that is}\,,~~T_1\equiv 0\,.\eeq{es2}
The corresponding equations are
\beq M\ddot{u}(t)+\Gb\dot{u}(t) =Q\cos \Go t\,,\n
 \Gr \ddot{w}(t)+\f{\pi^2}{4l^2}T_0 w(t)= 0\,.\eeq{es5}
We substitute the general solutions of these linear equations into the identity \eq{es2} and obtain the oscillating displacements with a constant (negative) shift of $u(t)$
 \beq u(t)=-\f{Q}{\Go(M^2 \Go^2+\Gb^2)}\gl(M\Go\cos\Go t -\Gb\sin\Go t+\sqrt{M^2 \Go^2+\Gb^2}\gr)\,,\n
 w(t)=A\cos \f{\Go t}{2} + B \sin \f{\Go t}{2}\,,~~~T_1(t)=0\,,\n
 A = \sqrt{\gl(\sqrt{M^2\Go^2+\Gb^2}+M\Go\gr)\psi}\,,~~~ B = \sqrt{\gl(\sqrt{M^2\Go^2+\Gb^2}-M\Go\gr)\psi}\,,\n
\psi= \f{16lQ}{\pi^2\Go(M^2\Go^2+\Gb^2)}\,,~~~(-u)_{max} = \f{2Q}{\Go\sqrt{M^2\Go^2+\Gb^2}}\,,\n
 w_{max}=\f{4}{\pi}\sqrt{\f{2lQ}{\Go\sqrt{M^2\Go^2+\Gb^2}}}=\f{4}{\pi}\sqrt{l (-u)_{max}}\,,~~~\Go = 2\Go_T
\,.\eeq{es6}

\begin{figure}[!ht]

\centering
\vspace*{0mm} \rotatebox{0}{\resizebox{!}{6.5cm}{%
\includegraphics [scale=0.25]{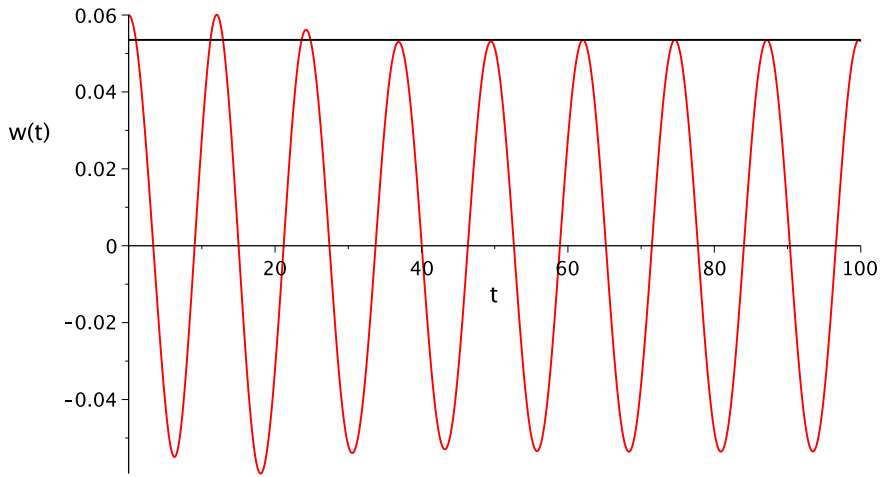}}}
 \caption{Forced oscillations. The transverse displacement $w(t)$.}
    \label{f4}
\end{figure}

\begin{figure}[!ht]

\centering
\vspace*{10mm} \rotatebox{0}{\resizebox{!}{6.5cm}{%
\includegraphics [scale=0.25]{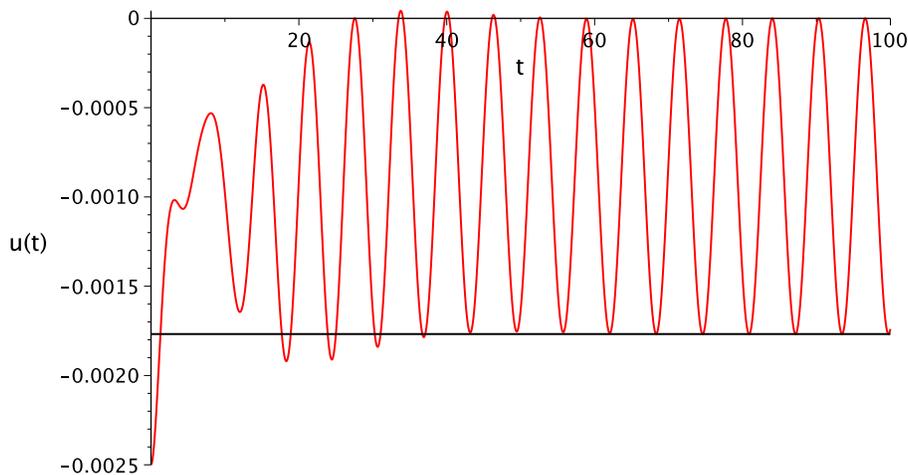}}}
 \caption{Forced oscillations. The longitudinal displacement $u(t)$.}
    \label{f5}
\end{figure}

\begin{figure}[!ht]

\centering
\vspace*{10mm} \rotatebox{0}{\resizebox{!}{6.5cm}{%
\includegraphics [scale=0.25]{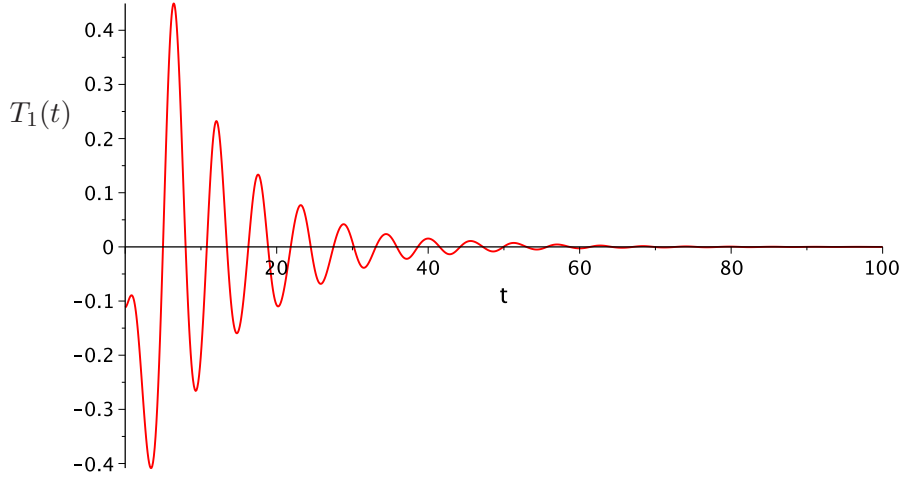}}}
\small\put(-340,137){$T_1(t)$}
 \caption{Forced oscillations. The function $T_1(t)$. It tends to zero in accordance with the analytical result for the established regime  \eq{es6}.}
    \label{f6}
\end{figure}

At least in a certain domain of the initial conditions, the transient solutions based on the equations  \eq{sm7} with $\Ga=0$ approach in time the above presented one based on the equations \eq{es2}, \eq{es5}. This is justified by numerical simulations of the transient problem with initial conditions inconsistent with the identity \eq{es2}. The numerically obtained oscillations and the oscillation amplitude found analytically from \eq{es2}, \eq{es5} are presented in \fig{f4}, \fig{f5} and \fig{f6}.  Remarkable, that the convergence to the establish regime occurs in spite of the fact that there is no dissipation directly associated with the lateral oscillations. The calculations were conducted  assuming $l=1$ m, $M=400$ kg, $\Gr=10$ kg/m, $T_0=1$ N, $Q=0.5$ N, $k=400$ N/m, $\Ga=0$, $\Gb=400$ Ns/m, $\Go=1$ 1/s, with the initial conditions $u(0)=-0.0025$ m, $w(0)=0.06$ m, $\dot{u}(0)=\dot{w}(0)=0$. The horizontal lines correspond to the amplitudes found analytically \eq{es6}.

\section{A more general analysis}
We now consider the dynamic equations \eq{sm7} with $\Gb=0$
\beq M\ddot{u}(t) + T_1(t) = Q\cos \Go t\,,\n
 \Gr \ddot{w}(t)+\Ga\dot{w}(t)+\f{\pi^2}{4l^2}T(t)w(t)= 0\,.\eeq{ga1}
We derive an approximate solution to these equations using harmonic analysis, where only main harmonics are preserved. So, for the steady oscillations we assume, neglecting higher harmonics, that the string oscillates transversally with frequency $\Go/2$
 \beq w(t)=A\cos\f{\Go t}{2} +B\sin\f{\Go t}{2}\eeq{ga2}
After substituting this in the expression for $T_1(t)$ in \eq{sm6}, we substitute $T_1(t)$  in the first equation of
\eq{ga1}. In doing so, we obtain and solve the equation with respect to $u(t)$. Finally, having expressions  for $w(t)$ and $u(t)$ we can express $T_1(t)$ explicitly. The latter two functions are found in the form
 \beq u(t) = -\f{\pi^2}{32 l}(A^2+B^2) + \f{1}{1-\Go_L^2}\gl[\gl(\f{Q}{k}-\f{\pi^2}{32 l}(A^2-B^2)\gr)\cos \Go t -\f{\pi^2}{16 l}AB\sin\Go t\gr],\n
T_1(t)=\f{1}{1-\Go_L^2}\gl[\gl(Q-\f{\pi^2k\Go_L^2}{32 l}(A^2-B^2)\gr)\cos\Go t-\f{\pi^2k\Go_L^2}{16 l}AB\sin\Go t\gr],~~~\Go_L=\f{\Go}{\GO_L}.\eeq{ga2t1}
Next, we substitute these expressions in the last equation in \eq{ga1}, which is satisfied on average with  $\cos \Go t$ and $\sin\Go t$ as the weights.
We obtain the following equations with respect to the coefficients
 \beq \f{\pi^2}{4l^2}A\gl[\f{1}{1-\Go_L^2}\gl(\f{Q}{2T_0}-\f{\pi^2k\Go_L^2}{64 l T_0}(A^2+B^2)\gr) + \gl(1-\f{\Go_T^2}{4}\gr)\gr]+\f{\Ga\Go}{2T_0}B=0\,,\n
        \f{\pi^2}{4l^2}B\gl[\f{1}{1-\Go_L^2}\gl(\f{Q}{2T_0}+\f{\pi^2k\Go_L^2}{64 l T_0}(A^2+B^2) \gr) - \gl(1-\f{\Go_T^2}{4}\gr)\gr]+\f{\Ga\Go}{2T_0}A=0\,,\n
        \Go_L = \Go/\GO_L\,,~~~\Go_T=\Go/\GO_T\,.\eeq{ga3}
From this the following solution is found
 \beq B= \Gg A\,,~~~ \Gg=-\Gk\gl(1-\sqrt{1-\f{1}{\Gk^2}}\gr)\,,~~~\Gk = \f{\pi^2Q}{4\Ga\Go l^2(1-\Go_L^2)}\eeq{ga4}
and
 \beq
 A=\pm\f{8l}{\pi\Go_L\sqrt{1+\Gg^2}}\sqrt{\f{Q}{2kl}\sqrt{1-\f{1}{\Gk^{2}}}+\f{T_0}{kl}\gl(1-\f{\Go_T^2}{4}\gr)(1-\Go_L^2)}
\,.\eeq{ga5}
It can be seen, in particular, that under the resonant excitation, $\Go_L=1$, the amplitude is independent of both the transverse oscillation frequency, $\GO_T$, and the viscosity level. Indeed, $\Gk \to \infty\,,~\Gg \to 0~~(\Go_L \to 1)$, and in the limit
 \beq  A=\pm\f{4l}{\pi}\sqrt{\f{2Q}{kl}}\,,~~B=0~~~(\Go_L=1)
\,.\eeq{ga6}
The solution \eq{ga4} - \eq{ga6} is valid in a domain, where the considered stationary oscillation regime really exists, that is, where it is stable. Below this issue is discussed in more detail.

Functions $u(t)$ and $T_1(t)$ for $\Go_L\ne 1$ are defined in \eq{ga2t1}. The limits at  $\Go_L= 1$ are
 \beq  u(t) = -\f{Q}{k} +\gl[\f{2T_0}{k}\gl(1-\f{\Go_T^2}{4}\gr) -\f{Q}{k}\gr]\cos \Go t + \f{4\Ga\Go l^2}{\pi^2k}\sin \Go t\,,\n
 T_1(t)= -2T_0\gl(1-\f{\Go_T^2}{4}\gr)\cos\Go t +\f{4\Ga\Go l^2}{\pi^2}\sin\Go t\,.\eeq{ga7}
If, in addition, $\Ga=0$ and $\Go_T=2~(\GO_T=\GO_L/2)$ then $T_1=0$, and we come back to the above-considered exact solution \eq{es6} (with $\Gb=0$).

\subsection{Energy flux}
In the above-considered general case with $\Gb=0$, the energy dissipation rate averaged over the period of oscillations is
 \beq D=2 \Ga\int_0^l \langle\dot{v}(x,t)^2\rangle\D x = \f{\Ga\Go^2l}{8}(A^2+B^2) = \f{\Ga\Go^2 l(1+\Gg^2)}{8} A^2=-\f{\Ga\Go^2 l \Gk\Gg}{4} A^2\,.\eeq{ef1}
On the other hand, it follows from \eq{ga2t1} that the energy flux produced by the external force is
 \beq N = 2Q \langle \dot{u}(t)\cos\Go t \rangle = -\f{\pi^2\Go\Gg}{16 l(1-\Go_L^2)}A^2Q\,.\eeq{ef2}
Referring to \eq{ga4} we see that these quantities coincide as it should be.

\section{Parametric resonance domains}
Neglecting nonlinear terms in equations \eq{ga3} we obtain an approximate expression for the main resonance domain boundary on the ($\Go,Q_L$) plane
 \beq Q= Q_L = \pm 2T_0(1-\Go_L^2)\sqrt{\gl(\f{2\Ga\Go l^2}{\pi^2T_0}\gr)^2 +\gl(1-\f{\Go_T^2}{4}\gr)^2}\,.\eeq{ef3}
Note that here $\Gk^2\ge 1$. In addition to this bound, the nonlinear analytical solution itself is bounded by the condition $T(t)\ge 0$, which is not satisfied if $\Gk^2<1$.

\subsection{Dimensionless formulation}
\begin{figure}[!ht]

\vspace{2mm}
\centering
\vspace*{2mm} \rotatebox{0}{\resizebox{!}{4.2cm}{%
\includegraphics [scale=0.25]{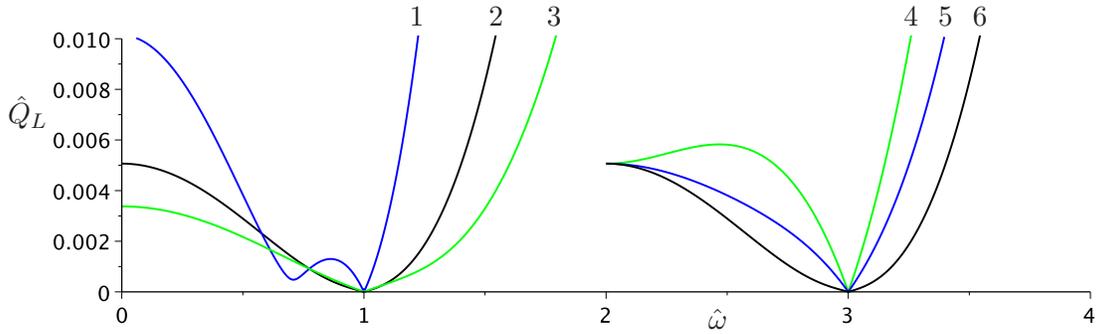}}}
\small \put(-150,2) {$\hat\Go$}
\small\put(-415,80){$\hat Q_L$}
\small\put(-263,117){1} \small\put(-233,117){2} \small\put(-211,117){3}
\small\put(-76,117){4} \small\put(-63,117){5} \small\put(-50,117){6}
 \caption{The PR domain based on the linear formulation. At the left: $\hat{Q}_L(\hat{\Go})$ for  $\Gm=0.025,\hat \Ga=2/15$ and $\Gl=\Gl_0\nu$, where  $\Gl_0=395$ corresponds to $\GO_T=1/2$. The curves correspond to $\nu=0.5 \, (1),  1\,  (2), 1.5 \, (3)$. At the right: $\hat Q_L(\hat\Go-2)$ for  $\Gm=0.025, \Gl=395$ and $\hat\Ga=(2/15)p$. The curves correspond to $p=10 \, (4), 5\, (5), 1\, (6)$.}
    \label{LINnu}
\end{figure}

We now introduce the dimensionless amplitude of $w(t)$, $\CA$,  and the other quantities
 \beq \hat \CA =\f{\CA}{l}=\f{A\sqrt{1+\Gg^2}}{l}\,,~~ (\hat{u},\hat{w})=\f{(u,w)}{l}\,,~~~\hat{t}=t\sqrt{\f{k}{M}}\,,~~~\hat{\Go}=\Go\sqrt{\f{M}{k}}\,,\n
 \hat{\Ga} =\f{\Ga}{\Gr}\sqrt{\f{M}{k}}\,,~~~\hat{\Gb}=\f{\Gb}{\sqrt{kM}}\,,~~~ (\hat{T}(t),\hat{Q})=\f{(T(t),Q)}{kl}\,,~~~\Gl=\f{kl}{T_0}\,,~~~\Gm=\f{\Gr l}{M}\,.\eeq{smn1}
In these terms,
 \beq \hat{\GO}_L=1\,,~~~\hat{\Go}_L=\hat{\Go}\,,~~~\hat{\GO}_T= \f{\pi}{2\sqrt{\Gl\Gm}}\,,~~~\hat{\Gk}=\f{\pi^2 \hat{Q}}{4\Gm(1-\hat{\Go}^2)\hat{\Ga}\hat{\Go}}
\,.\eeq{smnn2}
 The dynamic equations \eq{sm7} take the form (in the below relations, the hat symbol is omitted)
\beq \ddot{u}(t) +\Gb\dot{u}(t)+u(t) +\f{\pi^2}{16}w^2(t)=Q\cos\Go t\,,\n
 \ddot{w}(t)+\Ga\dot{w}(t)  +\GO_T^2\gl[1+\Gl\gl(u(t)+\f{\pi^2}{16}w^2(t)\gr)\gr]w(t)=0\eeq{smn2a}
and
 \beq   A=\pm\f{8}{\pi\Go\sqrt{1+\Gg^2}}\sqrt{\f{Q}{2}\sqrt{1-\f{1}{\Gk^2}}+\f{1}{\Gl}\gl(1-\f{\Go_T^2}{4}\gr)(1-\Go^2)}\,.\eeq{smnn3}
The $\Go - Q_L$ relation \eq{ef3} becomes
 \beq Q_L = \pm \f{2}{\Gl}|1-\Go^2|\sqrt{\gl(1-\f{\Go_T^2}{4}\gr)^2+\gl(\f{2}{\pi^2}\Gl\Gm\Ga\Go\gr)^2}\,.\eeq{ef3nd}
Some plots corresponding to this expression are presented in \fig{LINnu}.

 \subsection{Refinement of the bounds}
Now the following dimensionless parameters are fixed: $\hat\Ga=2/15, \hat\Gb=1/150, \Gm =0.025$. Also the expression for $Q$ is fixed: $\hat Q=1/(3\Gl)$ that corresponds to a third of the initial tensile force. The parameter $\Gl$ and frequency $\hat\Go$ are variable. We determine three regions. The first is the $L$-region of $\hat\Go$, where $Q_L<Q$, that is, where the PR exists in the linearized model in accordance with \eq{ef3nd}. Next is the $T$-region, where the $T_1$ amplitude found in \eq{ga2t1} is below the initial tensile force, $T_0$, that is, where the tensile force remains nonnegative and the above-obtained results can be acceptable. At last, the $R$-region is considered, where the steady oscillations arising under the variable non-negative tensile force are obtained numerically based on equations \eq{smn2a}. The regions are as follows
  \begin{tabbing}
 The casekkkkkkkkkkkkkkkk\= sssssssssssssssssssssss\=fffffffffffffffffffffffffffffffffffffff\= \kill \\
 ~~~~~~~~~~The case \> $~~~~~~~L$-region\> $~~~~~~~~~~T$-region\> ~~~~$R$-region\\
 $~~\Gl= 263.3$ ($\GO_T=\sqrt{3/8}$)\> $0.8362 <\hat\Go< 1.2883$\>~~~$0.8880 <\hat\Go<1.4631$\>$0.99<\hat\Go<1.3$\\
 $~~\Gl= 394.8$ ($\GO_T=1/2$)\> $0.7860 <\hat\Go< 1.1636$\>~~~$0.7348 <\hat\Go<1.1785$\>$0.85<\hat\Go<1.19$\\
 $~~\Gl= 526.4$ ($\GO_T=\sqrt{3}/4$)\> $0.7436<\hat\Go<1.1063$\>~~~$0.6456 <\hat\Go<1.0060$\>$0.77<\hat\Go<0.93$
 \end{tabbing}

It is remarkable that there is no dramatic difference between these regions. In addition, the regions, where the lateral oscillations exist, $PR$-regions, appear to be close to the corresponding $L$-regions with minor shifts towards higher frequencies. However, in each case, the $R$-region is only a part of the $PR$-region. The latter also includes the  irregular-oscillation regions adjacent to the $R$-region. In the $R$-regions, the steady oscillation amplitudes found numerically coincide with the analytical values with a good accuracy. Only the constant shift of the mass (see \eq{ga2t1}) can differ markedly in some cases (this could happen as a result of the neglect of the higher harmonics).

\begin{figure}[!ht]
\vspace{0mm}
\centering
\vspace*{0mm} \rotatebox{0}{\resizebox{!}{5cm}{%
\includegraphics [scale=0.35]{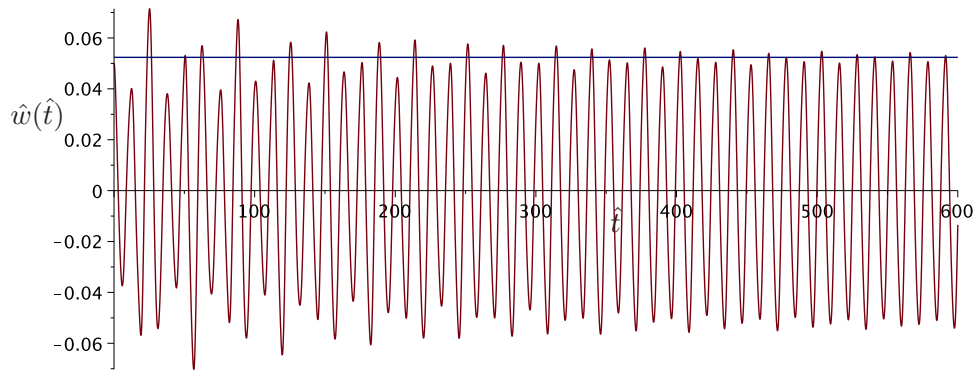}}}
\small \put(-367,96) {$\hat w(\hat t)$}
\put(-140,55){$\hat t$}

\vspace*{-2mm}
 \caption{Transient transverse oscillations: $\hat w(\hat t)$. }
    \label{f8}
\end{figure}

\begin{figure}[!ht]

\vspace{-3mm}
\centering
\vspace*{0mm} \rotatebox{0}{\resizebox{!}{5cm}{%
\includegraphics [scale=0.25]{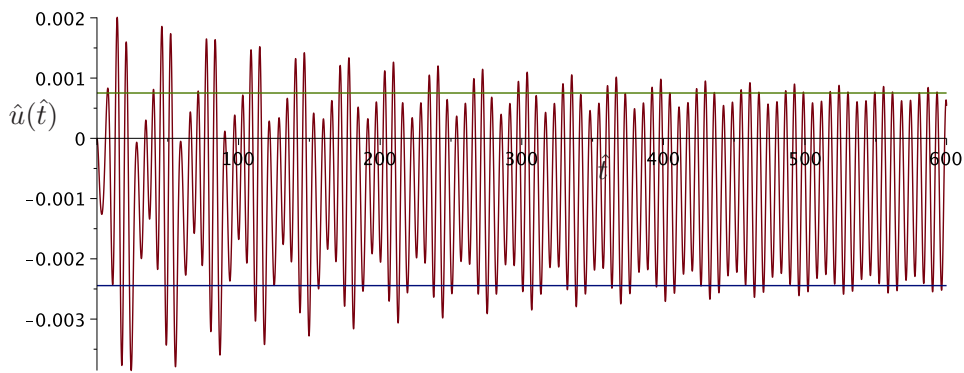}}}
\small \put(-367,96) { $\hat u(\hat t)$}
\put(-140.5,75){$\hat t$}

\vspace*{-2mm}
 \caption{Transient longitudinal oscillations: $\hat u(\hat t)$.}
    \label{f9}
\end{figure}

\begin{figure}[!ht]

\vspace{0mm}
\centering
\vspace*{0mm} \rotatebox{0}{\resizebox{!}{5cm}{%
\includegraphics [scale=0.25]{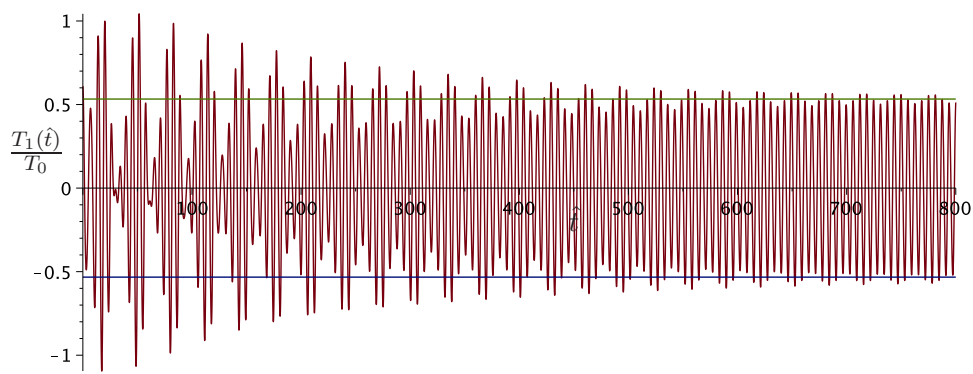}}}
\small \put(-367,83) {$\f{T_1(\hat t)}{T_0}$}
\put(-155,56){$ \hat t$}

\vspace*{-4mm}
 \caption{The $T_1(t)/T_0$-ratio. }
    \label{f10}
\end{figure}

\begin{figure}[!ht]
\centering
\vspace*{0mm} \rotatebox{0}{\resizebox{!}{5cm}{%
\includegraphics [scale=0.25]{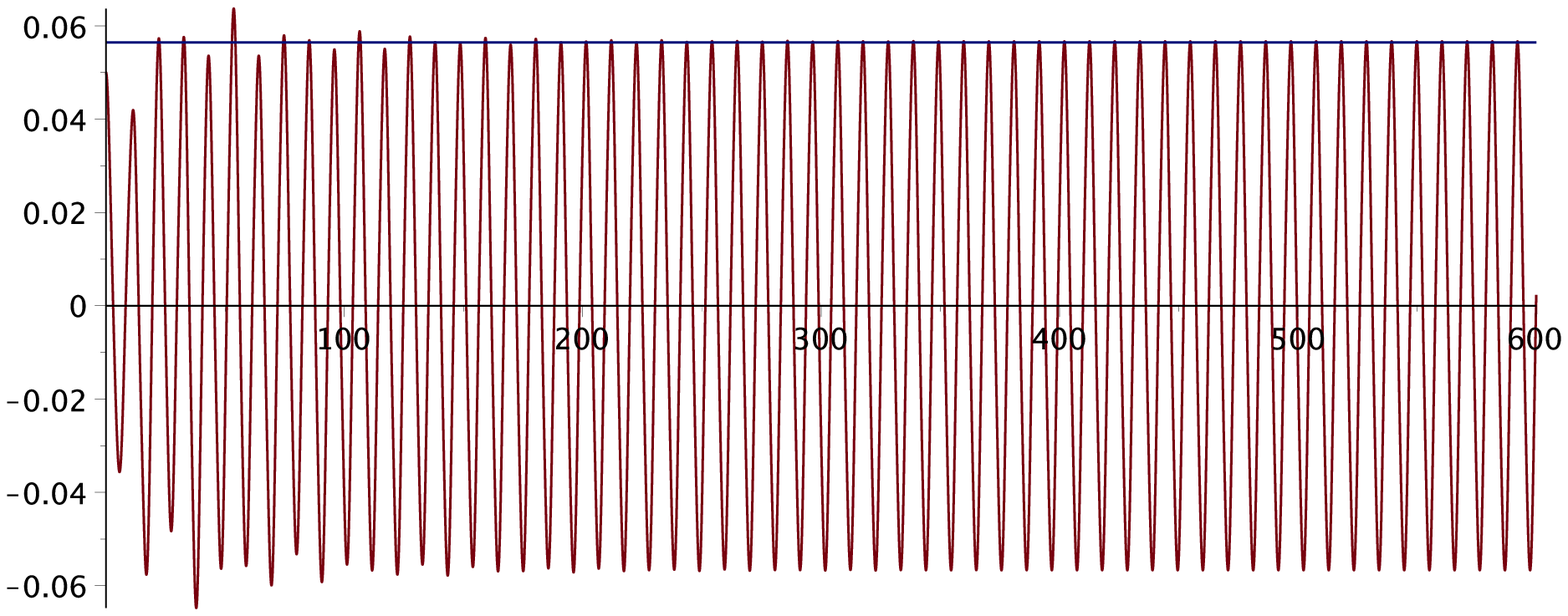}}}
\small \put(-374,100) {$\hat w(\hat t)$}
\put(-150,56){$\hat t$}
 \caption{Transient transverse oscillations on the upper bound of $R$-region: $\hat w(\hat t)$ for $\Gm=0.025, \hat\Ga=2/15, \GO_T=1/2, \hat\Go=1.19$.  In this case, the oscillation amplitudes quickly approach the  values found analytically.}
    \label{f11}
\end{figure}

\begin{figure}[!ht]

\vspace{0mm}
\centering
\vspace*{-3mm} \rotatebox{0}{\resizebox{!}{5cm}{%
\includegraphics [scale=0.25]{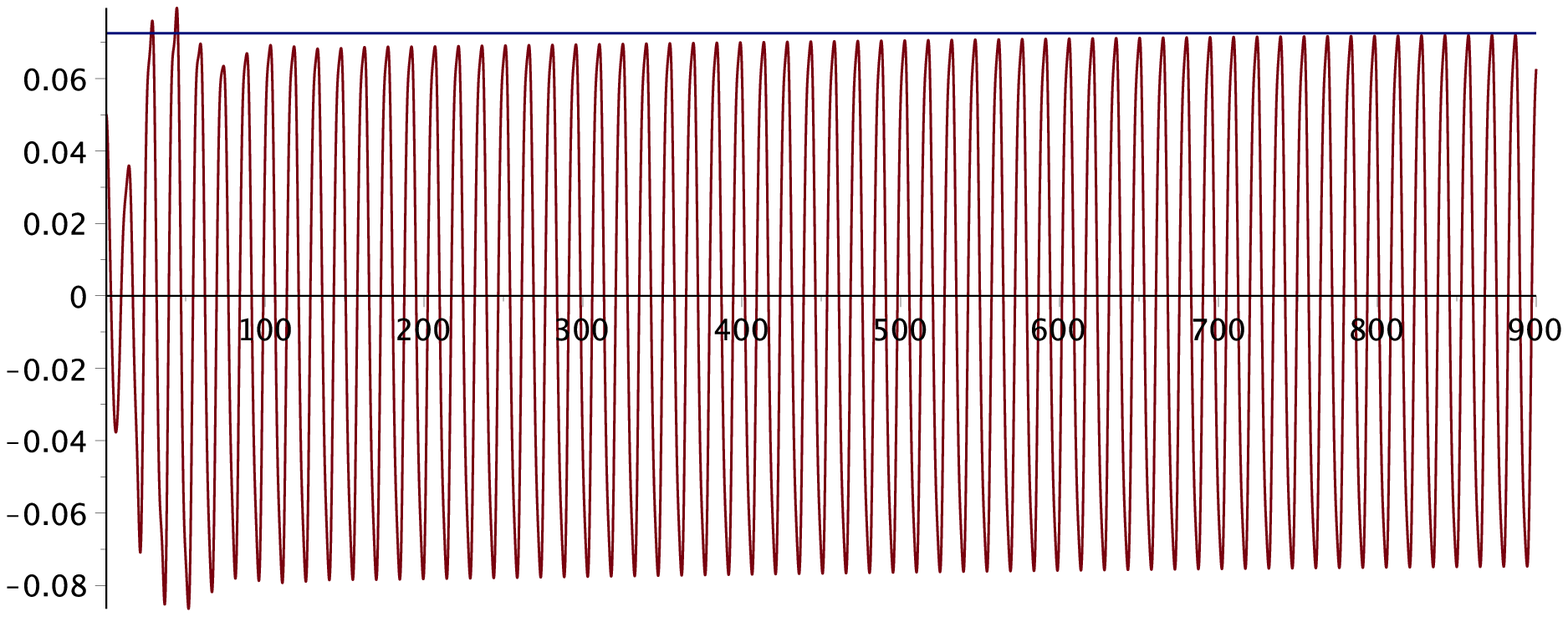}}}
\small \put(-374,96) {$\hat w(\hat t)$}
\put(-139,57){$\hat t$}
 \caption{Transient transverse oscillations on the lower bound of $R$-region: $\hat w(\hat t)$ for $\Gm=0.025, \hat\Ga=2/15, \GO_T=1/2, \hat\Go=0.85$.  In this case, the oscillation amplitudes slowly approach the  values found analytically.}
    \label{f12}
\end{figure}

\begin{figure}[!ht]

\vspace{2mm}
\centering
\vspace*{2mm} \rotatebox{0}{\resizebox{!}{4.5cm}{%
\includegraphics [scale=0.25]{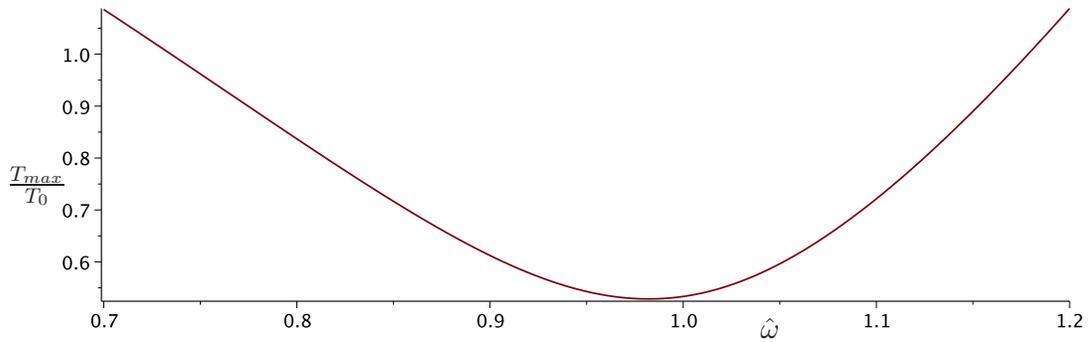}}}
\small\put(-410,55){$\f{T_{max}}{T_0}$}
\small\put(-125, 0){$\hat \Go$}
 \caption{The amplitude of $T_1(\hat t)/T_0$-ratio as a function of $\hat\Go$ found analytically for $\Gm=0.025, \hat\Ga=2/15, \GO_T=1/2$. Recall that the range of $\hat\Go$ is acceptable, where it is less than unity.}
    \label{f13}
\end{figure}

\begin{figure}[!ht]

\vspace{0mm}
\centering
\vspace*{0mm} \rotatebox{0}{\resizebox{!}{5cm}{%
\includegraphics [scale=0.25]{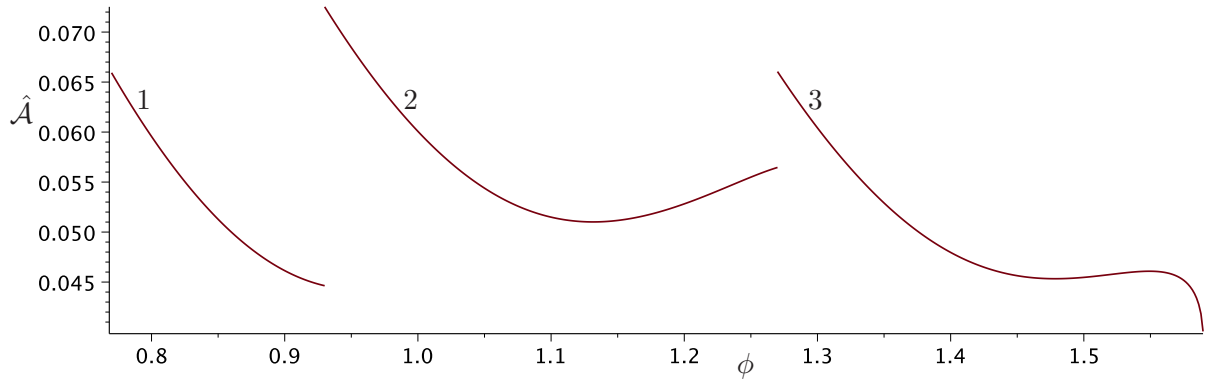}}}
\small\put(-407,100){1} \small\put(-306,100){2} \small\put(-153,100){3}
\small\put(-455,95){$\hat\CA$} \small\put(-180,0){$\Gf$}
 \caption{The analytically obtained amplitude of $\hat w(\hat t)$: $\hat\CA(\Go)$  for $\Gm=0.025, \hat\Ga=2/15$. The curves correspond to $\GO_T = \sqrt{3}/4$, $\hat\Go=\phi$ (1);  $\GO_T = 1/2$, $\hat\Go=\phi-0.08$ (2) and  $\GO_T = \sqrt{3/8}$, $\hat\Go=\phi-0.28$ (3).}
    \label{f14}
\end{figure}

Some illustrations are presented in the below figures. Transient transversal and longitudinal oscillations and the dynamic-to-static transient tensile force ratio for $\Gm=0.025, \hat\Ga=2/15, \GO_T=1/2, \hat\Go=1$ (in the $R$-region) are shown in \fig{f8}, \fig{f9}, and \fig{f10}, \res. Transient transversal oscillations for the border values of the frequency in the $R$-region, $\hat\Go=1.19$ and $\hat\Go=0.85$, under the same other conditions are presented in \fig{f11} and \fig{f12}, \res. In all these figures, the horizontal lines correspond to the amplitudes obtained analytically. In \fig{f13}, the amplitude of $T_1(\hat t)/T_0$-ratio is plotted as a function of $\hat\Go$ found analytically also under the same other conditions. Recall that the range of $\hat\Go$ is acceptable, in which this ratio does not exceed unity. Lastly, we show the amplitude-frequency characteristics, $\hat\CA(\hat\Go)$, with respect to the lateral oscillations for all three cases, \fig{f14}.

\vspace{1mm}

\clearpage

The right limiting point in \fig{f14}, where $\hat\CA= 0  \, (\hat\Go  \approx 1.31)$, limits the region where regular PR oscillations can be excited ($\hat\Gk^2<1$ for greater $\hat\Go$, and the solution \eq{ga5} failed).

\section{Conclusion}
The solutions obtained in this paper allow seeing how the system parameters affect the vibration level. For example, let the excitation frequency be resonant with respect to the lateral oscillations, $\Go_T=2$, and $\hat\Gk^2>1$. It can be seen from \eq{ga5} (also see \eq{sm2}, \eq{ga2t1} and \eq{ga4}) that the amplitude, $\CA$, increases unboundedly as the end mass, $M$, decreases.  Indeed, $\GO_L \sim 1/\sqrt{M}$ and $\CA\sim 1/\Go_L \sim 1/\sqrt{M}$. Note, however, that ``proportional" ($\sim$) remains asymptotically valid only for small amplitudes. In addition, the results allow to estimate the PR domains with steady oscillations and to determine the corresponding amplitude-frequency characteristics as was demonstrated in the paper.

In designing and setting a PR-based machine, a simple model used in this paper may be not sufficient, mainly because the interaction with the treated granular material is not included in the dynamic equations explicitly (only damping reflects this interaction). A properly developed model appears difficult for the analytical study. In this case, numerical simulations can be effective (as in designing of PR-based vibrating screen in LPMC), whereas the above-obtained results based on a simple model may be useful for preliminary estimations.

\vspace{5mm}
\noindent {\bf Acknowledgement.} This paper has been written while V.I. Slepyan held a short-term visiting research position at Tel Aviv University in the framework of the project ``VIBRO-IMPACT MACHINES BASED ON PARAMETRIC RESONANCE: Concepts, mathematical modelling, experimental verification and implementation", 01/01/2012-31/12/2015. We thankful for the support provided by FP7-People-2011-IAPP, Marie Curie Actions, Grant No. 284544, \\ http://www.openaire.eu/fr/component/openaire/project\_\,info/default/530?grant=284544 .

\newpage
\vskip 18pt
\begin{center}
{\bf  References}
\end{center}
\vskip 3pt
\noindent
[1]  V.I. Slepyan, I.G. Loginov and L.I. Slepyan, The method of resonance excitation of a vibrating sieve and the vibrating screen for its implementation. Ukrainian patent on invention No. 87369, 2009.

\noindent
[2] A.A. Vitt and G.S. Gorelik, Oscillations of an elastic pendulum as an example of the oscillations of two parametrically coupled linear systems.
 Journal of Technical Physics, 3 (1933)  294--307 (originally published in the Russian journal Zhurnal Tekhnicheskoy Fiziki, 3 (1933) 294--307).

 \noindent
[3] H.M. Lai, On the recurrence phenomenon of a resonant spring pendulum. Am. J. Phys. 52 (1984) 219-223.

 \noindent
[4] A.V. Gaponov-Grekhov, M.I. Rabinovich, Nonlinearities in action. Springer-Verlag, Berlin, London, 1992.

\noindent
[5] R. Baskaran and K. L. Turner, Mechanical domain coupled mode parametric resonance and amplification in a torsional mode micro electro
mechanical oscillator. J. Micromech. Microeng. 13 (2003) 701–707.

\noindent
[6] J.F. Rhoads, S.W. Shaw, K.L. Turner, R. Baskaran, Tunable microelectromechanical filters that exploit parametric resonance. Journal of Vibration and Acoustics 127 (2005) 423-430.

 \noindent
[7] S. Krylov, Parametric excitation and stabilization of electrostatically actuated microstructures. Int. J. Microscale Computational Engineering 6 (2008) 563-584.

 \noindent
[8] S. Krylov, I. Harari, and Y. Cohen, Stabilization of electrostatically actuated microstructures using parametric excitation. J. Micromech. Microeng 15 (2005) 1188–1204.

 \noindent
[9] S. Krylov, Y.Gerson, T. Nachmias,, and U. Keren, Excitation of large-amplitude parametric resonance by the mechanical stiffness
modulation of a microstructure. J. Micromech. Microeng 20 (2010) 1-12, 015041  doi:10.1088/0960-1317/20/1/015041.

 \noindent
[10] S. Krylov, N. Molinazzib, T. Shmilovicha, U. Pomerantza, S. Lulinskya, Parametric excitation of flexible vibrations of micro beams by fringing electrostatic fields. Proceedings of the ASME 2010 International Design Engineering Technical Conferences \& Computers and Information in Engineering Conference IDETC/CIE 2010. Montreal, Quebec, Canada, 601-611, 2010.

 \noindent
[11] Fey, R.H.B., Mallon, N.J., Kraaij, C.S., and Nijmeijer, H.,  Nonlinear resonances in an axially excited beam carrying a top mass : simulations and experiments. Nonlinear Dynamics, 66(3) (2011), 285-302.

 \noindent
[12] Fossen, T.I., and Nijmeijer, H., Parametric resonance in dynamical systems, Springer, New York, 2012.

 \noindent
[13]  H.Plat, I. Bucher, Optimizing parametric oscillators with tunable boundary conditions. J. of Sound and Vibration, 332 (2013) 487–493.

\end{document}